\documentclass[12pt]{article}
\usepackage{subfigure}
\usepackage{latexsym}
\usepackage{amsmath}
\usepackage{amssymb}
\usepackage{amsthm}
\usepackage{undertilde}
\usepackage{graphicx}
\usepackage{hyperref}
\usepackage[latin1]{inputenc}
\usepackage[T1]{fontenc}
\usepackage[all]{xy}
\makeatletter
\renewcommand{\@seccntformat}[1]
{\csname the#1\endcsname.\enspace} \makeatother
\def \XXint#1#2#3{{\setbox0=\hbox{$#1{#2#3}{\int}$}
     \vcenter{\hbox{$#2#3$}}\kern-.5\wd0}}

\begin{document}
\newtheorem{theorem}{Theorem}
\newtheorem{lemma}{Lemma}
\newtheorem{remark}{Remark}
\newtheorem{remarks}{Remarks}
\newtheorem{corollary}{Corollary}
\newtheorem{example}{Example}
\newtheorem{definition}{Definition}
\newtheorem{proposition}{Proposition}
\newcommand{\R}{\mathbb{R}}
\newcommand{\sgn}{\mbox{sgn}}
\thispagestyle{empty}
\begin{center}
{\bf\Large A Weighted Model Confidence Set: Applications to
Local and Mixture Model Confidence Sets}\\
Amir.T. Payandeh Najafabadi\footnote{Corresponding author.
Email: amirtpayandeh@sbu.ac.ir}, Ghobad Barmalzan, \& Shahla Aghaei\\
Department of Mathematical Sciences, Shahid Beheshti University,
G.C. Evin, 1983963113, Tehran, Iran\\
 \today
\end{center}
\begin{abstract}
This article provides a weighted model confidence set, whenever
underling model has been misspecified and some part of support of
random variable $X$ conveys some important information about
underling true model. Application of such weighted model
confidence set for local and mixture model confidence sets have
been given. Two simulation studies have been conducted to show
practical application of our findings.
\end{abstract}
\section{Introduction}
Approximating an unknown density function  $h(\cdot)$ is an
interesting problem which has a wide range of applications in
statistical inference and data analysis. To find out an
appropriate approximation, in first step, one has to collect a
collection of family of distributions, say models, which can be
appropriate (in some sense) competing for $h(\cdot).$ Then, in the
second step, unknown parameters of such selected models have to be
estimated using an appropriated (in some sense) estimation method.
Such collection of appropriate competing models \emph{either}
belong to a family of distributions (say  a class of nested
models) \emph{or} a collection $k(\geq 2)$ families of
distributions (say a class of non-nested models). In the most
statistical approaches using visual (such as qq--plots) or
nonparametric (e.g. Kolmogorov--Smirnov test) tools, a nested
models has been selected. Then, using an appropriate estimation
method (such maximum likelihood, Bayesian, etc) unknown density
$h(\cdot)$ has been approximated. But for the non-nested approach,
selecting an appropriate non-nested models is a difficult task.

A considerable body of literature has been devoted to inference
under a class of non-nested models. For instance: using the
generalized likelihood ratio (${\cal LR}$) test, Cox (1961, 1962)
developed a statistic test to compare two non-nested models.
Atkinson (1970) considered a combined models as an appropriated
competing for an unknown model. Then, he derived a statistical
testing procedure to: ({\bf 1}) study departure from one model in
the direction of another and ({\bf 2}) test the hypothesis that
all fitted models are equivalent. Pesaran (1974) employed the
Cox's test (developed for comparing separate families of
hypotheses) to study the choice between two non-nested linear
single-equation econometric models. Pesaran \& Deaton (1978)
extended Pesaran (1974)'s findings to cover multivariate nonlinear
models whenever full information maximum likelihood estimation is
available. Davidson \&  MaKinnson (1981, 2002) considered several
popular procedures which test an econometric model and they
established those procedures are closely related, but not
identical, to the non-nested hypothesis tests which was proposed
by Pesaran \& Deaton (1978). Pesaran (1987) emphasized the
distinction between a ``local null'' and a ``local alternative''.
Then, under local alternatives, they derived the asymptotic
distribution of the Cox's test statistic. Fisher \& McAleer (1981)
derived two modified ${\cal LR}$ tests that are asymptotically
equivalent to the Cox's test. Dastoor (1983) studied different
between Cox's and Atkinson's statistics. In 1989, using concept of
the Kullback--Leibler divergence, Vuong extended Cox's test. The
Vuong's test compares two competing density functions based upon
the expectation of their ${\cal LR}$ statistic. Based upon
minimization of expectation of the Akaike Information Criteria
(AIC), Shimodaira (1998) constructed a confidence set. Lo, et al.
(2001) showed that under Vuong's assumptions the ${\cal LR}$
statistic based upon the Kullback--Leibler divergence and random
sample $X_1,X_2,\cdots,X_n$ for hypothesis test
\begin{center}
${\cal H}_0: X_i\sim$ $k_0$-component normal mixture~~~v.s.~~${\cal H}_1:
X_i\sim$ $k_1$-component normal mixture
\end{center}
is asymptotically distributed as a weighted sum of independent
chi-squared random variables with one degree of freedom. Hansen,
et al. (2003), via a simulation study, showed that the model
confidence set captures the superior models across a range of
significance levels. Lu, et al. (2008) compared the Wald's and
Cox's statistics based upon a simulation study. They showed that
the small sample behavior of both statistics is closed to their
asymptotic distributions. Hansen, et al. (2011) employed
confidence set approach to study density of inflation in a given
data set and the best Taylor rule regression for an empirical
problem. Sayyareh (2012) constructed a tracking interval for the
problem of selecting an appropriate estimation of unknown density
$h(\cdot)$ among a class of non-nested families of distributions
whenever data arrived from a Type II right censored phenomena. Li,
et al. (2012) provided a Bayesian approach to the problem of
nonparametric estimation of unknown density $h(\cdot).$

Using the Kullback--Leibler (${\cal KL}$) divergence along with
Vuong's test, this article constructs a set of appropriate
weighted models, say weighted model confidence set, for unknown
true density $h(\cdot)$ which is a subset of a class of non-nested
models. A weighted confidence set is a set of models that is
constructed such that it will contain the best model with a given
level of confidence.  All models which take to the weighted model
confidence set are equivalent in terms of closeness to true
density $h(\cdot).$ Applications of such weighted model confidence
set has been given for local and mixture goodness of fitness.
Practical implementation of the results has been confirmed through
two simulation studies.

The rest of this article developed as follows. Section 2 collects
some pertinent concepts of the Kullback--Leibler divergence,
likelihood ratio tests and some mathematical background for the
problem. Main results have been explored in Section 3. Two
simulation studies have been conducted in Section 4.
\section{Preliminaries}
The Kullback--Leibler divergence, say ${\cal KL},$ is widely
employed statistic to select an appropriate model among a class of
competing models. Suppose $\log \frac{h(X)}{f(X; \theta)}$ stands
for the log--likelihood ratio for density function $f(\cdot)$ and
unknown true density $h(\cdot)$, for an observation $X.$ The
expectation of this ratio, with respect to true density
$h(\cdot),$ is ${\cal KL}$ divergence between $f(\cdot)$ and
$h(\cdot)$. In other words,
\begin{eqnarray*}
{\cal KL}(h|| f)&=& E_{h} \left\{\log \left(\frac{h(X)}{f(X;
\theta) }
\right)\right\}\\
&=& E_{h}\{ \log h(X)\} - E_{h}\left\{\log f(X; \theta)\right\},
\end{eqnarray*}
where $ E_{h}\{\log h(X)\}$ and $ E_{h}\left\{\log f(X;
\theta)\right\}$ present irrelevant and relevant parts of ${\cal
KL}$ divergence, respectively. The expectation $E_{h}\{\log f(X;
\theta)\}$  plays a vital role in this article. The ${\cal KL}$
divergence is a nonnegative value which cannot be consider as a
distance. But from its definition, one may conclude that ${\cal
KL}(h|| f) = 0$ implies that $h(\cdot)=f(\cdot; \theta),$ for all
$\theta\in\Theta.$ To select an appropriate set of models among a
class of non-nested competing models, it suffices to consider just
relevant part of the ${\cal KL}$ divergence.

The Hellinger and $L^2$ distances between two density functions
$f$ and $g$ have been defined by
\begin{eqnarray}
  \label{Hellinger-L2-distance}\nonumber H(f;g)&=&\sqrt{\int_{\Bbb
R}\left(\sqrt{f(x)}-\sqrt{g(x)}\right)^2dx};\\
 L^2(f;g)&=&\sqrt{\int_{\Bbb R}(f(x)-g(x))^2dx}.
\end{eqnarray}
The Hellinger and $L^2$ are two symmetric and non-negative
distances that satisfies the triangle inequality. Moreover,
convergence in Kullback--Leibler divergence implies convergence of
the Hellinger and  $L^2$ distances, see Van Erven \& Harremos
(2014) for more details. On other hand, the Kullback--Leibler
divergence has a probabilistic/statistical meaning while Hellinger
and $L^2$ distances do not. Therefore, in model selection
literature the Kullback--Leibler divergence is a well-known
distance which measures information lost whenever $f$ is used to
approximate $g,$ see Lv \& Liu (2014) for more details.

Suppose $X_1, X_2,\cdots, X_n$ is a sequence of i.i.d. random
variables with common and unknown density function $h(\cdot)$.
Moreover, suppose that two competing classes of models
\begin{eqnarray*}
  {\cal F}_{\Theta} &=& \left\{f(\cdot; \theta):~ \theta \in \Theta
\nonumber \subseteq {\Bbb R}^{p} \right\} \\
 {\cal G}_{\Gamma} &=& \left\{ g(\cdot; \gamma):~
 \gamma\in \Gamma \subseteq {\Bbb R}^{q}\right\}
\end{eqnarray*}
can be viewed as two appropriate approximations for $h(\cdot),$
where $\Theta$ and $\Gamma,$ respectively, represent two different
parameter spaces with dimensions $p$ and $q$. Two classes of
models ${\cal F}_{\Theta}$ and ${\cal G}_{\Gamma}$ are called two
non-nested classes if and only if ${\cal F}_{\Theta}\cap {\cal
G}_{\Gamma}=\emptyset.$ Note that, the model ${\cal G}_{\Gamma}$
is nested in ${\cal F}_{\Theta}$ if and only if $ {\cal
G}_{\Gamma} \subset {\cal F}_{\Theta}.$  The following represents
formal definition of non-nested models which employs in model
selection literature. The model ${\cal F}_{\Theta}$ with respect
to $h(\cdot)$ is called well-specified if and only if there exists
$\theta\in \Theta$ such that $h(\cdot)\equiv f(\cdot; \theta),$
Otherwise, model ${\cal F}_{\Theta}$ with respect to $h(\cdot)$ is
misspecified.

The most popular criteria in model selection has been coined and
developed by Fisher (1921, 1922). In his seminal work, he employed
concept of the maximum likelihood estimation method to develop a
selection method criteria, well-known as a maximum likelihood
method. This article utilizes a more advanced version of the
maximum likelihood method, well-known as a quasi maximum
likelihood estimator, say QMLE, that will be used in the rest of
this article. The following represents definition of the QMLE.
\begin{definition}
Suppose $X_1, X_2, \cdots, X_n$ is a random sample with unknown
density $h(\cdot).$ Moreover, suppose that weighted density
function $f(x; \theta)$ is an appropriate (in some sense)
approximation for $h(\cdot).$ An estimator
\begin{eqnarray*}
{\widehat \theta}^{n}=\arg\sup_{{\theta} \in \Theta} \,\sum
_{i=1}^{n} \log f(X_i; \theta)
\end{eqnarray*}
which is maximized log--likelihood of ${\cal L}^{f}_{n}(\theta)=
\sum _{i=1}^{n} \log f(X_i; \theta)\}$ is well-known as the QMLE
 for parameter $\theta.$  Moreover, an estimator
\begin{eqnarray*}
\theta^{\star}= \arg \, \max_{\theta \in \Theta} E_{h}
\left\{\frac{1}{n}\,\,{\cal L}^{f}_{n}(\theta)\right\}.
\end{eqnarray*}
which minimizes the $\cal KL$ divergence is called pseudo
estimator for $\theta.$
\end{definition}
It is well-known that the usual MLE coincides with the QMLE
whenever the family ${\cal F}_{\Theta}$ is well-specified with
respect to the $h(\cdot)$, see White (1982b) for more details.
Huber (1976) showed that the QML estimator,
${\widehat{\theta}}^{n}$, is a consistent estimator for pseudo
estimator $\theta^{\star}$ whenever ${\cal F}_{\Theta}$ is
misspecified with respect to $h(\cdot)$.

Since $E_{h}\{\log h(X)\}$ is free of $\theta,$  therefore based
upon random sample $X_1,\cdots, X_n,$ two approximations
$f_\theta(\cdot)$ and $g_\gamma(\cdot)$ for $h(\cdot)$ can be
compared through their relevant parts of $\cal KL$ divergence.

Now, we collect some useful lemmas relating to the empirical
distributions and likelihood ratio statistics which are useful in
representing the main results of this article. The following from
Barmalzan \& Payandeh Najafabadi (2012) provides an ML estimator
for unknown density function $h(\cdot).$
\begin{lemma} (Barmalzan \& Payandeh Najafabadi, 2012)
\label{two_steps} Suppose $X_1,\cdots, X_n$ is a sequence of
i.i.d. random variables with common and unknown density function
$h(\cdot).$ Moreover, suppose that for given constant $t$ random
variable $Y_{i}(t)$ has been defined by $Y_i(t)=I(X_i \leq t)$ for
$i=1, \cdots, n$. Then, $Y_{1}(t),\cdots,Y_{n}(t)$ are i.i.d.
random variables with common Bernoulli distribution with unknown
parameter $H(t)= P(X_1 \leq t).$ Therefore,
\begin{description}
    \item[i)] The ML estimator for $H(t)$ is
        \begin{eqnarray*}
\widehat{H}(t)= \frac{1}{n} \sum_{i=1}^{n} Y_i(t)= \frac{1}{n}
\sum_{i=1}^{n}I(X_i \leq t);
\end{eqnarray*}
    \item[ii)] The ML estimator for $h(t),$ based upon sample $x_1,\cdots, x_n,$ is
        $\widehat{h}(x_{i})=1/n,$ for $i=1,\cdots,n;$
    \item[iii)] The ML estimator for $E_{h}\{\log  w(X; \theta_0) \, f(X; \theta_1)\},$ based upon sample $x_1,\cdots, x_n,$ is
            \begin{eqnarray*}
E_{\widehat{h}}\{\log  f(X;{\widehat{\theta}^n)}\}
&=&\frac{1}{n}\sum_{i=1}^{n} \log f(x_i; {\widehat{\theta}^n});
            \end{eqnarray*}
    \item[iv)] An estimator $ E_{\widehat{h}}\{\log  f(X; {\widehat{\theta}^n})$ is converged, in
    probability, to $E_{h}\{\log f(X; \theta)\}.$
\end{description}
\end{lemma}
Based upon a random sample $X_1,\cdots, X_n$ from unknown true
density $h(\cdot),$ the likelihood ratio statistic, ${\cal LR}$,
for ${\cal F}_{\Theta}$ against ${\cal G}_{\Gamma}$ is denoted by
${\cal LR}_{n}({\widehat \theta}^{n}, \, {\widehat \gamma}^{n})$
and defined by subtracting log--likelihood of two family of
weighted models \eqref{weighted_models}, i.e.,
\begin{eqnarray*}
{\cal LR}_{n}({\widehat \theta}^{n}, \, {\widehat
\gamma}_{n})&=&{\cal L}^{f}_{n}({\widehat \theta}^{n}) - {\cal
L}^{g}_{n}({\widehat \gamma}^{n}),
\end{eqnarray*}
where $f\in{\cal F}_{\Theta}$ and $g\in{\cal G}_{\Gamma}.$

The following recalls some useful properties of ${\cal
LR}_{n}({\widehat \theta}^{n}, \, {\widehat \gamma}^{n}),$ proof
may be found in White (1982a) and Vuong (1989).
\begin{lemma}
\label{Likelihood_ratio_statistics} Suppose $X_1,\cdots, X_n$ be a
random sample with common and unknown true density $h(\cdot).$
Then, the likelihood ratio statistic ${\cal LR}_{n}({\widehat
\theta}^{n}, \, {\widehat \gamma}^{n})$ has the following
properties:
\begin{description}
    \item[(i)] Under some mild conditions (White, 1982a) ${\cal LR}_{n} ^{w}  \,({\widehat \theta}^{n}, \, {\widehat
\gamma}^{n})/n$ converges, almost surly, to
 \begin{eqnarray*}
        { E_{h} \left\{ \log\,
\frac{f(X; {{ \theta}^{\star}})}{g(X; {{\gamma}^{\star}})}
\right\}};
\end{eqnarray*}
    \item[(ii)] If $f(\cdot; {\theta^{\star}})\neq g(\cdot; {\gamma^{\star}})$,
    then $n^{- \,1/2}\, {\cal LR}_{n}  \,({\widehat \theta}^{n}, \, {\widehat
\gamma}^{n}) - n^{ \,1/2}B_{\star}$ converges, in law, to normal
distribution $N(0, \, A^{2}_{\star}),$
\end{description}
where
\begin{eqnarray*} A^{2}_{\star}= Var _{h}\left\{\log\,
\frac{f(X;{{ \theta}^{\star}} )}{ g(X;{{\gamma}^{\star}} )
}\right\},~~~\&~~B_{\star}=E_{h} \left\{\log\,  \frac{f(X; {{
\theta}^{\star}} )}{ g(X; {{\gamma}^{\star}}) } \right\}
\end{eqnarray*} and its estimator is
\begin{eqnarray*}
\widehat {A}^{2}_n= \frac{1}{n}\, \sum_{i=1}^{n} \left\{\log\,
\frac{f(X_i; {{\widehat \theta}^{n}})}{g(X_i; {{\widehat
\gamma}^{n}})} \right\}^2 - \left\{\frac{1}{n}\sum_{i=1}^{n}
\log\, \frac{f(X_i; {{\widehat \theta}^{n}})}{g(X_i; {{\widehat
\gamma}^{n}})}\right\}^2.
\end{eqnarray*}
\end{lemma}
The following section provides a model confidence set for true
density function based upon a collection of weighted non-nested
models, say weighted model confidence set. Application of such
weighted model confidence set for local and mixture confidence
sets have been given in two subsections.
\section{Weighted Model Confidence Set}
In traditional model-fitting strategy whole of observations
received equal weight. Inference based on naive use of whole of
information may be erroneous, since information conveyed by some
of the observations is less important (in some sense) than the
information conveyed by others. On the other hand, in many
applications, only some parts of the space of variables are of
interest, so that it makes sense to focus our attention in those
regions. Weighted distribution are ideally suited model for these
phenomenons. The weight function can be determined (or selected)
to {\it either} reflecting such facts {\it or} taking into account
some related information.

Suppose $\delta(x;\theta_0)$ is a nonnegative function  with
finite expectation. Then, weighted density function
$f^w(x;\theta_0,\theta),$ based on realization $x$ of random
variable $X$ under density function $f(x;\theta_1),$ has been
defined by
\begin{eqnarray}
\label{weighted-density-function}
  f^w(x;\theta_0,\theta) &=&
  \frac{\delta(x;\theta_0)}{E_{f}(\delta(X;\theta_0))}f(x,\theta)\\
\nonumber  &=& w(x;\theta_0)f(x,\theta)
\end{eqnarray}
where $x\in{\Bbb R}.$ Since increasing number of unknown
parameters contradict with the parsimony's principle (Posada \&
Buckley, 2004) and artificially impact on the Kullback--Leibler
divergence (Barmalzan \& Sayyareh, 2010). Hereafter now, we just
consider a situation that {\it either} $\theta_0$ is a given
constant {\it or} $\theta_0=\theta.$ Therefore, we just have
$f^w(x;\theta),$ where $\theta\in\Theta$ and $x\in{\Bbb R}.$

Fisher (1934) developed concept of weighted distribution. Rao
(1965) pointed out that: in many situations the recorded
observations cannot be considered as a random sample from the
original distribution due to non-observability of some events,
damage caused to original observations, etc. The length biased
distribution (weighted distribution with $w(x;\theta_0)=x/E_h(X)$)
has been found various applications in biomedical areas such as
early detection of a disease. breast cancer (Zelen \& Feinleib,
1969), human families and wild-life population studies (Rao,
1965), cardiology study involving two phases (Cnaan, 1985).

Weighted model selection has not received much attention in the
model selection literature. The most of existence researches have
been done regarding to bayes factor (Larose \& Dey, 1996, 1998),
nested model selection (Cheung, 2005, Ingrassia, et al., 2014),
weighted model selection criteria, such as weighted least-squares
support vector machines (Cawley, 2006), some  information
criterion, such as BIC and ICL, for mixture models as a weighted
model (Dang, et al., 2014).

Now, Suppose that there is a collection of $k$ competing weighted
non-nested models which could be used to describe random sample
$X=(X_1,\cdots, X_n)$ obtained under common and unknown density
$h(\cdot).$
\begin{eqnarray}
\label{weighted_models}  {\cal F}^{w}_{\Theta(i)} &=& \left\{
f^w(\cdot; \theta(i)):~ \theta_i \in \Theta(i)  \subseteq {\Bbb
R}^{p(i)} \right\}
\end{eqnarray}
where $\Theta(i)$ is parameter space with dimensions $p(i)$ and
$w(\cdot)$ is a nonnegative and given weight function. Moreover,
suppose that $\cal U$ denotes a collection of $k(\geq2)$ weighted
non-nested family of models ${\cal F}^{w}_{\Theta(i)},$ for $i\in
{\cal M}=\{1,2,\cdots,k\},$ i.e., ${\cal U}^w=\bigcup_{i\in{\cal
M}}{\cal F}^{w}_{\Theta(i)}.$

In the traditional model selection, the Kullback--Leibler
divergence, ${\cal KL},$ is equally penalized all support of
random variable $X.$ Weighted version of the Kullback--Leibler
divergence can be defined as
\begin{eqnarray*}
{\cal KL}\,\{h(X)|| f^{w}(X; \theta)\}&=& E_{h}\{ \log h(X)\} -
E_{h}\left\{\log f^{w}(X; \theta)\right\}.
\end{eqnarray*}
Now using the relevant expectation $E_{h}(\log f^{w}(X; \theta))$
one may conclude that: The class of models ${\cal
F}^{w}_{\Theta(i)}$ can be considered as an appropriate
approximation for unknown density $h(\cdot)$ if and only if the
null hypothesis ${\cal H}_{0i}$ in the following hypothesis test
\begin{eqnarray}
\label{Hypothesis_test}
\left\{ {\begin{array}{*{20}c}
   { {\cal H}_{0i}:E_{h}\{\log f^{w}(X; {\theta^{\star}(i)})\}
\geq max _{j  \,\in {\cal M} \backslash \{i\}  } \, E_{h}\{\log f^{w}(X; {\theta^{\star}(j)})\},     } & {} & { }  \\
   {} & {} & {}  \\
   { {\cal H}_{1i}:E_{h}\{\log f^{w}(X; {\theta^{\star}(i)})\}
< \ max _{j  \,\in {\cal M} \backslash \{i\}  }\, E_{h}\{\log f^{w}(X; {\theta^{\star}(j)})\},  } & {} & { }  \\
\end{array}} \right.
\end{eqnarray}
has been accepted at significance level $\alpha,$ where
$\theta^{\star}(i)= \arg \, \max_{\theta \in \Theta(i)} E_{h}
\{\frac{1}{n}\,\,{\cal L}^{f^{w}}_{n}(\theta(i))\}.$

The null hypothesis ${\cal H}_{0i}$, for $i\in {\cal M}=\{1,
\cdots, n\}$ in hypothesis test (\ref{Hypothesis_test}), at
significance level $\alpha,$ will be rejected  in favor of ${\cal
H}_{1i}$ if and only if
\begin{eqnarray*}
\left\{ \min_{{j  \,\in {\cal M}} \backslash \{i\}   } \, T_{ij}<
-z_{\frac{\alpha}{k-1}}\right\},
\end{eqnarray*}
where
\begin{eqnarray*}
T_{ij}:=\frac{n^{- \,1/2} \left\{{\cal L}_{n}^{f^{w}}(\widehat
{\theta}^{\star}(i)) - {\cal L}_{n}^{f^{w}}
(\widehat{\theta}^{\star}(j))- (\dim(\theta(i))- \dim(\theta(j)))
\right\}}{{{\widehat A}_{n}}},
\end{eqnarray*}
$\dim(\cdot)$ denotes dimension and $z_{p}$ is the $p^{\hbox{th}}$
quantile of standard normal distribution, and ${\cal L}^{f^w}$
stands for the log--likelihood based upon random sample
$X_1,\cdots, X_n,$ i.e., ${\cal L}^{f^w}_{n}(\theta)= \sum
_{i=1}^{n} \log \{ w(X_i; \theta) \,f(X_i; \theta)\}.$ For more
details about this hypothesis test in the non-nested models and
its applications, interested readers may refer to Barmalzan \&
Payandeh Najafabadi (2012), among others.

To simplify the idea which behind of our weighted confidence set,
consider a situation that we have only two competing weighted
non-nested family of models  ${\cal F}^{w}_{\Theta(1)}$ and ${\cal
F}^{w}_{\Theta(2)}.$ One may setup following two hypothesis tests
\footnotesize{\begin{eqnarray*} {\cal H}_{0}^{(1)}:E_{h}\{\log
f^{w}(X; {\theta_{\star}(1)})\} \geq E_{h}\{\log
f^{w}(X;{\theta_{\star}(2)} )\} &v.s.&
{\cal H}_{1}^{(1)}:E_{h}\{\log f^{w}(X; {\theta_{\star}(1)})\} < E_{h}\{\log f^{w}(X; {\theta_{\star}(2)})\};\\
and\\
{\cal H}_{0}^{(2)}:E_{h}\{\log f^{w}(X; {\theta_{\star}(2)})\} \geq
E_{h}\{\log f^{w}(X; {\theta_{\star}(1)})\} &v.s.&
{\cal H}_{1}^{(2)}:E_{h}\{\log f^{w}(X; {\theta_{\star}(2)})\} < E_{h}\{\log
f^{w}(X;{\theta_{\star}(1)} )\}.
\end{eqnarray*}}\normalsize
Suppose that ${\bf\tau}$ denotes a $100(1-\alpha)\%$ weighted
model confidence set for $h(\cdot).$ Using the above two
hypothesis tests, one may include ${\cal F}^{w}_{\Theta(1)}$ in
${\bf\tau}$ whenever ${\cal H}_{0}^{(1)}$ has been accepted at
significance level $\alpha/(2-1)=\alpha.$  Similarly,  one may
include ${\cal F}^{w}_{\Theta(2)}$  or both ${\cal
F}^{w}_{\Theta(1)}, {\cal F}^{w}_{\Theta(2)}$ in ${\bf\tau}$
whenever ${\cal H}_{0}^{(2)}$ and  ${\cal H}_{0}^{(1)}$, ${\cal
H}_{0}^{(2)}$ has been accepted at significance level $\alpha,$
respectively. Therefore, the $100(1-\alpha)\%$  weighted model
confidence set is one of these three sets ${\bf\tau}=\{{\cal
F}^{w}_{\Theta(1)}\},$ ${\bf\tau}=\{{\cal F}^{w}_{\Theta(2)}\},$
or ${\bf\tau}=\{{\cal F}^{w}_{\Theta(1)}; {\cal
F}^{w}_{\Theta(2)}\}.$

The following theorem formalizes the above idea. Its proof is
similar to  Barmalzan \& Payandeh Najafabadi (2012, Theorem 1).
\begin{theorem}
\label{confidence-set} Suppose $X_1, X_2,\cdots, X_n$ is a random
sample with common and unknown density $h(\cdot).$ Moreover,
suppose that $\cal U$ is a collection of $k(\geq2)$ weighted
non-nested family of models ${\cal F}^{w}_{\Theta(i)},$ for $i\in
{\cal M}:=\{1,2,\cdots,k\}.$ Then, the $100(1-\alpha)\%$  weighted
model confidence set for  $h(\cdot)$ is given by
\begin{eqnarray*}
  \tau^{w} &=& \displaystyle\cup_{i\in{\cal M}}\left\{{\cal F}^{w}_{\Theta(i)}:~{\cal
F}^{w}_{\Theta(i)}\in{\cal U}^w~\&~\min_{{j \,\in {\cal M}}
\backslash \{i\} } \, T_{ij}> c\right\},
\end{eqnarray*}
where $c:=z_{\alpha/(k-1)}.$
\end{theorem}
Note that the above weighted model confidence set is not empty
because it at least contains the maximum $E_{h} \{\log w(X;
\theta) f(X; \theta)\}$ model with an error smaller than
significance level $\alpha.$
\begin{remark}
Suppose $X_1, X_2,\cdots, X_n$ is a random sample with common
unknown true density $h(\cdot).$ Moreover, suppose $\cal U^\omega$
is a collection of $k(\geq2)$ weighted non-nested models
\begin{eqnarray*}
  {\cal F}_{\Theta(i)}^\omega &=& \left\{f^{w}(\cdot; {\theta(i)}):=\frac{\delta(\cdot; \theta(i))}{E_h(\delta(X; \theta(i)))}f_{\theta(i)}(\cdot):~ \theta(i) \in \Theta(i) \subseteq {\Bbb
R}^{p(i)} \right\},~~i\in {\cal M}=\{1,2,\cdots,k\}.
\end{eqnarray*}
Then, the $100(1-\alpha)\%$ weighted model confidence set for
unknown true density $h(\cdot)$ is given by
\begin{eqnarray*}
  \tau^w &=& \displaystyle\cup_{i\in{\cal M}}\left\{{\cal F}^{w}_{\Theta(i)}:~{\cal
F}_{\Theta(i)}^\omega\in{\cal U}^{w}~\&~\min_{{j \,\in {\cal M}}
\backslash \{i\} } \, T_{ij}^\omega>
z_{\frac{\alpha}{k-1}}\right\},
\end{eqnarray*}
where
\begin{eqnarray*}
T_{ij}^\omega:=\frac{n^{- \,1/2} \left\{{\cal
L}_{n}^{f_{i}^\omega}(\widehat {\theta}^{\star}(i)) - {\cal
L}_{n}^{f_{j}^\omega} (\widehat{\theta}^{\star}(j))-
(\dim(\theta(i))- \dim(\theta(j))) \right\}}{{{\widehat
A}_{n}^\omega}}.
\end{eqnarray*}
\end{remark}
The weight function $w(\cdot; \theta)$ selects based upon nature
of problem in the hand or local goodness of fitness. The above
idea may also develop to mixture model selection.
\subsection{Application to local Model Confidence Set}
In practical situations, sometimes different parts of the data (or
support of random variable $X$) may be weighted differently. This
local consideration is perfectly reasonable, because it provides
lower variance for observations and consequently more information,
see Hand \& Vinciotti (2003) for more details. On the other hand,
in many practical problems all parts of the distributions are not
conveyed of equal information about under study phenomenon. In the
situation where the true model is properly specified, this fact is
not a big issue since under these circumstances a good fit in some
region will not lead a poor fit in another region. However, the
problem arrives whenever model is misspecified that a good fit in
some region may well reduce from quality of fit in another region.
Improving the fit of a misspecified model in some specified part
of the space by forcing a close fit between the model and the
underlying distributions in that region may be achieved through
differential weighting. Note, however, that the choice of weights
ignores the fact that different points may be of differing degree
of importance in the context of the problem.

Using result of Theorem \ref{confidence-set}, the following
provides a local model confidence set.
\begin{remark}
\label{Local_goodness} Suppose $X_1, X_2,\cdots, X_n$ is a random
sample with common and unknown true density $h(\cdot).$ Moreover,
suppose that $A$ is a subset of support $X$ where conveys some
important information. Then, the $100(1-\alpha)\%$ model
confidence set in subset $A,$ say local model confidence set, is
given by:
\begin{eqnarray*}
\displaystyle  \tau^{w} &=& \displaystyle\bigcup_{i\in{\cal
M}}\left\{{\cal F}_{\Theta(i)}^{w}:~{\cal
F}_{\Theta(i)}^{w}\in{\cal U}^{w}~\&~\min_{{j \,\in {\cal M}}
\backslash \{i\} } \, T_{ij}^{w}> z_{\frac{\alpha}{k-1}}\right\}.
\end{eqnarray*}
where ${\cal U}^w$ is a collections of $k(\geq2)$ weighted
non-nested models, $w(x)=I_A(x)/P_h(X\in A),$  and $I_A(\cdot)$
stands for the indicator function.

It is worthwhile mentioning that the above set $\tau^{w}$ provides
a $100(1-\alpha)\%$ confidence set for true density $h(\cdot)$
just on subset $A\subset{\Bbb R}$ not for whole of support $X.$
\end{remark}
\subsection{Application to Mixture Model Confidence Set}
Using weighted model confidence set, one may go beyond of regular
and classical model confidence set and consider mixture model
confidence set. Such $100(1-\alpha)\%$ mixture confidence set can
be obtained by the following three steps:
\begin{description}
    \item[Step 1:] Partition whole of support of random variable $X$ into
    $m$ partitions;
    \item[Step 2:] Construct a $100(1-\beta)\%,$ where $0<\beta\leq 1-(1-\alpha)^{1/m},$ local confidence set for each
    partition;
    \item[Step 3:] Estimate an optimal mixture weight using a different criteria.
\end{description}
Step 3 provides an optimal $100(1-\alpha)\%$ mixture confidence
set from another optimal criteria.

The following theorem illustrates the above three steps whenever
support of random variable $X$ partitions into two disjoint sets
$A$ and $A^\prime,$ i.e., $A\cup {A^\prime}={\Bbb R}$ and $A\cap
{A^\prime}=\varnothing.$
\begin{theorem}
\label{confidence_set_2-mixture} Suppose $X_1, X_2,\cdots, X_n$ is
a random sample with common and unknown true density $h(\cdot)$
and ${\cal U}^{w_1}$ and ${\cal W}^{w_2}$  are two collections of
$k_1(\geq2)$ and $k_2(\geq2)$ non-nested models, respectively,
which locally appropriated for $h(\cdot).$ Moreover, suppose that
\normalsize{\begin{eqnarray*} \displaystyle  \tau^{w_1} &=&
\displaystyle\bigcup_{i\in{\cal M}_1}\left\{{\cal
F}_{\Theta(i)}^{w_1}:~{\cal F}_{\Theta(i)}^{w_1}\in{\cal
U}^{w_1}~\&~\min_{{j \,\in {\cal M}_1} \backslash \{i\} } \,
T_{ij}^{w_1}>
z_{\frac{\beta}{k_1-1}}\right\};  \qquad {\cal {M}}_{1} \in \{1, 2,\cdots, k_1\}\\
\displaystyle  \tau^{\omega_2} &=& \displaystyle\bigcup_{i\in{\cal
M}_1}\left\{{\cal G}_{\Gamma(i)}^{w_2}:~{\cal
G}_{\Gamma(i)}^{w_2}\in{\cal W}^{w_2}~\&~\min_{{j \,\in {\cal
M}_2} \backslash \{i\} } \, T_{ij}^{w_2}>
z_{\frac{\beta}{k_2-1}}\right\}; \qquad {\cal {M}}_{2} \in \{1,
2,\cdots, k_2\}
\end{eqnarray*}}\normalsize
are two $100(1-\beta)\%$ local model confidence set for unknown
density $h(\cdot)$ where $0<\beta<1-\sqrt{1-\alpha}.$ Then, an
optimal $100(1-\alpha)\%$ global model confidence set for unknown
density $h(\cdot)$ which minimized ${\cal KL}$ distance between
convex combination of elements of $\tau^{w_1}$ and $\tau^{w_2}$
and $h(\cdot)$ is given by \footnotesize{\begin{eqnarray*}
  \tau &=& \displaystyle\bigcup_{(i,j)\in{\cal M}_1\times{\cal M}_2}\left\{\alpha^{opt}_{ij}{f}^{w_1}(x;{{\hat\theta}(i)} )+(1-\alpha^{opt}_{ij}){g}^{w_2}(x; {{\hat\gamma}(j)}):~~{f}^{w_1}(x;{{\hat\theta}(i)}) \in
\tau^{w_1}~\&~{g}^{w_2}(x; {{\hat\gamma}(j)}) \in \tau^{w_2}
\right\},
\end{eqnarray*}}\normalsize
where $w_1(x)=I_A(x)/P_h(X\in A)$ and
$w_2(x)=I_{A^\prime}(x)/P_h(X\in {A^\prime})$ are two given weight
functions in two partitions $A$ and ${A^\prime}$ (where $A\cup
{A^\prime}={\Bbb R}$ and $A\cap {A^\prime}=\varnothing$),
\begin{eqnarray}
\label{alpha_Opt}
\displaystyle\alpha^{opt}_{ij}&=&\min\{1,~\max\{0,~argmax\{{\hat\Psi}(\alpha_{ij})\}\}\},
\end{eqnarray}
and ${\hat\Psi}(\alpha_{ij})
=\sum_{l=1}^{n}\left(\log(\alpha_{ij}{f}^{w_1}(x_l;
{{\hat\theta}(i)})+(1-\alpha_{ij}){g}^{w_2}(x_l;{{\hat\gamma}(j)}
))\right)/n.$
\end{theorem}
{\bf Proof.} Suppose $f(\cdot):=\alpha_{ij}{f}^{w_1}(\cdot;
{{\hat\theta}(i)})+(1-\alpha_{ij}){g}^{w_2}(\cdot;
{{\hat\gamma}(j)} )$ is a member of $\tau.$ Therefore,
\begin{eqnarray*}
P(f\in \tau)&=&P\left(\alpha_{ij}{f}^{w_1}(x;{{\hat\theta}(i)} )+(1-\alpha_{ij}){g}^{w_2}(x; {{\hat\gamma}(j)}) \in \tau\right)\\
&=&P\left({f}^{w_1}(x;{{\hat\theta}(i)} ) \in \tau^{w_1}\, and \, {g}^{w_2}(x; {{\hat\gamma}(j)}) \in \tau^{w_2}\right)\\
&=& P\left({f}^{w_1}(x;{{\hat\theta}(i)} ) \in \tau^{w_1}\right)\, P\left({g}^{w_2}(x; {{\hat\gamma}(j)}) \in \tau^{w_2}\right)\\
&=& P\left(\min_{{j \,\in {\cal M}_1} \backslash \{i\} } \,
T_{ij}^{w_1}> z_{\frac{\beta}{k_1-1}}\right)\,P\left(\min_{{j
\,\in {\cal M}_2} \backslash \{i\} } \, T_{ij}^{w_2}>
z_{\frac{\beta}{k_2-1}}\right) \\
&\geq &(1-\beta)\,(1-\beta)\\
&\geq&1-\alpha.
\end{eqnarray*}
On the other hand, $f(\cdot)$ is a convex combination of elements
of $\tau^{w_1}$ and $\tau^{w_2}.$ Now observe that the ${\cal KL}$
distance between $f(\cdot)$ and $h(\cdot)$ minimized if and only
if
\begin{eqnarray*}
  \Psi(\alpha_{ij}) &=& E_{h}\left(\log(\alpha_{ij}{f}^{w_1}(X; {{\hat\theta}(i)} )+(1-\alpha_{ij}){g}^{w_2}(X; _{{\hat\gamma}(j)}))   \right)
\end{eqnarray*}
has been maximized  in $\alpha_{ij}.$ The desired proof arrived
from the fact that
$\partial^2\Psi(\alpha_{ij})/\partial\alpha_{ij}^2\geq0$ and
${\hat\Psi}(\cdot)$ is an estimator for $\Psi(\cdot).$ $\square$
\section{Simulation Study}
This section through two simulation studies shows that how one may
employ the above findings in practical applications.
\begin{example}
Suppose random sample $x_1,x_2,\cdots,x_n$ (for $n=50,200, 300$)
have been generated from a length biased (i.e., $w(x)=x/x/E_h(X)$)
Lognormal model with parameters $\theta=(\mu=2,\sigma^2=0.5)$. For
this simulation, we consider the following three length biased
non-nested models.
\begin{eqnarray*}
{\cal F}_{\Theta(1)}&=&\left\{f(x; {\theta(1)} )=\frac{x}{exp\{\mu+ \frac{1}{2} \sigma^2\}}.\,
\frac{1}{x\,\sqrt{2 \pi \sigma^2}}\, \exp\left\{ -\frac{\left(\log
x -\mu\right)^2}{2\sigma^2}\right\};\quad \theta(1)=(\mu,
\sigma^2)\right\};\\
{\cal F}_{\Theta(2)}&=&\left\{f(x; {\theta(2)})= \frac{x}{\alpha\,\lambda}.\, \frac{1}{
\lambda^{\alpha}\, \Gamma(\alpha) }\, x^{\alpha -1} \,
\exp\left\{- \frac{x}{\lambda}\right\} ;\quad \theta(2)=(\alpha,
\, \lambda) \right\};~\hbox{and}\\
{\cal F}_{\Theta(3)}&=&\left\{f(x; {\theta(3)})= \frac{x}{\gamma \,\Gamma(1+\frac{1}{\beta})}.\,
\frac{\beta}{\gamma} \, \left(\frac{x}{\gamma}\right)^{\beta -1}
\, \exp\left\{ - \left(\frac{x}{\gamma}\right)^\beta\right\}
;\quad \theta(3)=(\beta, \, \gamma)\right\}.
\end{eqnarray*}
as a possible competing models to determine the underling density
$h(\cdot).$
\end{example}
To obtain a $95\%$ confidence set for the true density function
Lognormal $(2,0.5),$ using the Monte-Carlo method with the R
software, we generate, 1000 times, three sample size with length
$n=50,200,300$ from underling Lognormal model. Using such
generated data along with the above three non-nested models, to
build up a $95\%$ confidence set for true model $h(\cdot),$ one
has to conduct the following hypothesis test at significance level
$\alpha=0.05.$
\begin{eqnarray}
\label{hypothesis-tests} \nonumber{\cal H}_{01}: E_{h}\{\log
f(X;{\hat{\theta}(1)} )\} &\geq & \max_{j \in \{2,\, 3\}} \,
E_{h}\{\log
f(X;{\hat{\theta}(j)})\}~\hbox{v.s.}~ {\cal H}_{11}: Reject~{\cal H}_{01}; \\
{\cal H}_{02}: E_{h}\{\log f(X; {\hat{\theta}(2)})\} &\geq& \max_{j
\in \{1,\, 3\}} \, E_{h}\{\log
f(X;{\hat{\theta}(j)})\}~\hbox{v.s.}~ {\cal H}_{12}: : Reject~{\cal H}_{02};\\
\nonumber{\cal H}_{03}: E_{h}\{\log f(X; {\hat{\theta}(3)})\} &\geq
& \max_{j \in \{1,\,2\}} \, E_{h}\{\log
f(X;{\hat{\theta}(j)} )\}~\hbox{v.s.}~ {\cal H}_{13}:: Reject~{\cal
H}_{03}.
\end{eqnarray}
Table 1 shows decision of the above hypothesis tests at
significance level $\alpha=0.05$ for sample size $n=50,200,300.$
\begin{center}
\scriptsize{Table 1: Results of the above hypothesis tests at
level $\alpha=0.05.$} \tiny{
\begin{tabular}{c c c c c c c}
     \hline
      &\multicolumn{6}{c}{Hypothesis Test}\\
     \cline{2-7}
    & \multicolumn{2}{c}{${\cal H}_{01}$ v.s.  ${\cal H}_{11}$}& \multicolumn{2}{c}{${\cal H}_{02}$ v.s.  ${\cal H}_{12}$} & \multicolumn{2}{c}{${\cal H}_{03}$ v.s.  ${\cal H}_{13}$}\\
      \cline{2-7}
Sample Size   &  Test Statistic    & Conclusion     &  Test Statistic     & Conclusion     & Test Statistic & Conclusion \\
      \cline{1-7}
$n=50$   & 0.98 & ${\cal H}_{01}$ is accepted & -\,1.12 & ${\cal H}_{02}$ is accepted& -\,1.48 & ${\cal H}_{03}$ accepted \\
$n=200$  & 1.87 & ${\cal H}_{01}$ is accepted & -\,1.88 & ${\cal H}_{02}$ is accepted & -\,2.37 & ${\cal H}_{03}$ is rejected \\
$n=300$ & 2.24 & ${\cal H}_{01}$ is accepted& -\,2.24& ${\cal H}_{02}$ is rejected & -\,2.80 & ${\cal H}_{03}$ is rejected \\
      \hline
\end{tabular}}
\end{center}
Using results of Table 1,  the following $95\%$ confidence set for
the true density function Lognormal $(2,0.5)$ has been given in
Table 2.
\begin{center}
\scriptsize{Table 2: A $95\%$ confidence set for different sample
size}\tiny{\\
\begin{tabular}{ c c}
  \hline
  Sample size & A $95\%$ confidence set \\
  \hline
  $n=50$ & $\tau=\left\{ {\cal F}_{\Theta(1)};\, {\cal
F}_{\Theta(2)}; \, {\cal F}_{\Theta(3)}\right\}$ \\
  $n=200$ & $\tau=\left\{ {\cal F}_{\Theta(1)};\, {\cal F}_{\Theta(2)}\right\}$ \\
  $n=300$ & $\tau=\left\{ {\cal F}_{\Theta(1)}\right\}$ \\
  \hline
\end{tabular}}
\end{center}
The $95\%$ confidence set, given by Table 2, shows that the method
works properly and the true model (Lognormal) falls in confidence
set. For the interpretation of equivalence of the above confidence
sets, see Sayyareh, et al. (2011).

The following example explores a situation where one single
confidence set cannot consider as an appropriate model confidence
set for true density function.
\begin{example}
Suppose random sample $x_1,x_2,\cdots,x_{1000}$ have been
generated from a following density function.
\begin{eqnarray}
\label{True-density-Example-2} h(x)=\frac{1}{3}\,
\frac{1}{0.5}\, e^{-|x+4|/0.5}    + \frac{2}{3}\, \frac{
e^{-(x-6)} }    {(1+e^{-(x-6)})^2}, \qquad x \in{\Bbb R}.
\end{eqnarray}
A histogram and true density function have been illustrated in
part (a) of Figure 1.
\end{example}
From illustrated histogram, one may readily conclude that the
underling distribution $h(\cdot)$ is a continuous distribution
with two different modes. Thus, we cannot apply only one single
density on the support $X$. In this case, in the first step, we
divide whole of support into two partitions $A=(-\infty, 0]$ and
$A^\prime=(0, +\infty)$. Based upon graphical investigation, for
the first part, we consider the non-nested competing models:
\{normal, cauchy, Logistic and Laplace\} models and for the second
part we propose three non-nested competing models: \{gamma,
Weibull and lognormal\} models. Now in the following three steps,
we develop a $95\%$ mixture confidence set for such generated
observation.
\begin{description}
    \item[Step 2-1: Local Confidence Set for A.] In this partition, we consider the following four non-nested
models.
\begin{eqnarray*}
{\cal F}^{w_1}_{\Theta(1)}&=&\left\{f^{w_1}(x;{\theta(1)} )=w_1(x)
\frac{1}{\sqrt{2 \pi \sigma^2}}\, \exp\left\{ -\frac{\left( x
-\mu\right)^2}{2\sigma^2}\right\};\quad \theta(1)=(\mu,
\sigma^2), \, x \in {\Bbb R} \right\};\\
{\cal F}^{w_1}_{\Theta(2)}&=&\left\{f^{w_1}(x; {\theta(2)})=w_1(x)
\frac{1} { \nu \pi \{1+(\frac{x-\theta}{\nu})^2\} }; \quad
\theta(2)=(\theta,
\nu), \, x \in {\Bbb R}  \right\};\\
{\cal F}^{w_1}_{\Theta(3)}&=&\left\{f^{w_1}(x; {\theta(3)})=w_1(x)
\frac{ \frac{1}{\beta} \,e^{-(x-\gamma)/\beta} }
{(1+e^{-(x-\gamma)/\beta})^2}; \quad \theta(3)=(\gamma,
\beta), \, x \in {\Bbb R} \right\};\\
{\cal F}^{w_1}_{\Theta(4)}&=&\left\{f^{w_1}(x; {\theta(4)}
)=w_1(x) \frac{1}{b}\, e^{-|x-\lambda|/b}; \quad
\theta(4)=(\lambda, b), \, x \in {\Bbb R} \right\},
\end{eqnarray*}
where $w_1(x)=I_A(x)/P_h(X\in A).$ To build up a $975\%$
confidence set for first part, one has to conduct the following
hypothesis test at significance level $\beta=0.025.$
\begin{eqnarray*}
\label{hypothesis-tests} \nonumber{\cal H}^{f^{w_1}}_{01}:
E_{h}\{\log f^{w_1}(X; {\hat{\theta}(1)})\} &\geq & \max_{j \in
\{2,\,3,\,4\}} \, E_{h}\{\log
f^{w_1}(X;{\hat{\theta}(j)} )\}\quad~\hbox{v.s.}~\quad{\cal H}^{f^{w_1}}_{11}: Reject~{\cal H}^{f^{w_1}}_{01}; \\
{\cal H}^{f^{w_1}}_{02}: E_{h}\{\log f^{w_1}(X;
{\hat{\theta}(2)})\} &\geq& \max_{j \in \{1,\,3,\, 4\}} \,
E_{h}\{\log
f^{w_1}(X; {\hat{\theta}(j)})\}\quad~\hbox{v.s.}~\quad {\cal H}^{f^{w_1}}_{12}: Reject~{\cal H}^{f^{w_1}}_{02};\\
\nonumber{\cal H}^{f^{w_1}}_{03}: E_{h}\{\log f^{w_1}(X;
{\hat{\theta}(3)})\} &\geq & \max_{j \in \{1,\,2,\, 4\}} \,
E_{h}\{\log f^{w_1}(X;{\hat{\theta}(j)} )\}\quad~\hbox{v.s.}~\quad
{\cal H}^{f^{w_1}}_{13}: Reject~{\cal
H}^{f^{w_1}}_{03};\\
\nonumber{\cal H}^{f^{w_1}}_{04}: E_{h}\{\log
f^{w_1}(X;{\hat{\theta}(4)} )\} &\geq & \max_{j \in \{1,\,2,\,3\}}
\, E_{h}\{\log f^{w_1}(X;{\hat{\theta}(j)}
)\}\quad~\hbox{v.s.}~\quad {\cal H}^{f^{w_1}}_{14}: Reject~{\cal
H}^{f^{w_1}}_{04}.
\end{eqnarray*}
Table 3 shows decision of the above hypothesis tests at
significance level $\beta=0.025$ after 1000 irritations.
\begin{center}
\scriptsize{Table 3: Hypothesis tests at significant level
$\beta=0.025$ for
$A=(-\infty, 0]$} \tiny{\\
\begin{tabular}{c c c}\hline
Hypothesis Test & Test Statistics & Conclusion\\
\hline ${\cal H}^{f^{w_1}}_{01}$ v.s.  ${\cal H}^{f^{w_1}}_{11}$& $-2.609$& ${\cal H}^{f^{w_1}}_{01}$ is rejected\\
 ${\cal H}^{f^{w_1}}_{02}$ v.s.  ${\cal H}^{f^{w_1}}_{12}$& $-4.099$& ${\cal H}^{f^{w_1}}_{02}$ is rejected\\
 ${\cal H}^{f^{w_1}}_{03}$ v.s.  ${\cal H}^{f^{w_1}}_{13}$& $-1.210$& ${\cal H}^{f^{w_1}}_{03}$ is accepted\\
 ${\cal H}^{f^{w_1}}_{04}$ v.s.  ${\cal H}^{f^{w_1}}_{14}$& $0.0490$& ${\cal H}^{f^{w_1}}_{04}$ is accepted\\
\hline
\end{tabular}}
\end{center}
Therefore, the desired a $975\%$ local confidence set for the
first part is
\begin{eqnarray*}
\tau_1^{w_1}=\left\{ {\cal F}^{w_1}_{\Theta(3)}; {\cal
F}^{w_1}_{\Theta(4)} \right\}.\end{eqnarray*}
    \item[Step 2-2: Local Confidence Set for $A^\prime$.] For the second part, we consider the following three non-nested
competing models
\begin{eqnarray*}
{\cal G}^{w_2}_{\Gamma(1)}&=&\left\{g^{w_2}(x; {\gamma(1)})=w_2(x)
\frac{1}{ \lambda^{\alpha}\, \Gamma(\alpha) }\, x^{\alpha -1} \,
\exp\left\{- \frac{x}{\lambda}\right\} ;\quad \gamma(1)=(\alpha,
\, \lambda),\, x \in {\Bbb R}^{+} \right\};\\
{\cal G}^{w_2}_{\Gamma(2)}&=&\left\{g^{w_2}(x;{\gamma(2)} )=w_2(x)
\frac{\beta}{\gamma} \, \left(\frac{x}{\gamma}\right)^{\beta -1}
\, \exp\left\{ - \left(\frac{x}{\gamma}\right)^\beta\right\}
;\quad \gamma(2)=(\beta, \, \gamma),\, x \in {\Bbb R}^{+} \right\}\\
{\cal G}^{w_2}_{\Gamma(3)}&=&\left\{g^{w_2}(x; {\gamma(3)})=w_2(x)
\frac{1}{x\,\sqrt{2 \pi \sigma^2}}\, \exp\left\{ -\frac{\left(\log
x -\mu\right)^2}{2\sigma^2}\right\};\quad \gamma(3)=(\mu,
\sigma^2),\, x \in {\Bbb R}^{+} \right\},
\end{eqnarray*}
where $w_2(x)=I_{A^\prime}(x)/P_h(X\in A^\prime).$ Using the above
three competing models along with the following three hypothesis
tests,
\begin{eqnarray*}
\label{hypothesis-tests} \nonumber{\cal H}^{g^{w_2}}_{01}:
E_{h}\{\log g^{w_2}(X; {\hat{\gamma}(1)})\} \geq \max_{j \in
\{2,\, 3\}} \, E_{h}\{\log
g^{w_2}(X; {\hat{\gamma}(j)})\}~\hbox{v.s.}~ {\cal H}^{g^{w_2}}_{11}: Reject~{\cal H}^{g^{w_2}}_{01}; \\
{\cal H}^{g^{w_2}}_{02}: E_{h}\{\log g^{w_2}(X;{\hat{\gamma}(2)}
)\} \geq \max_{j \in \{1,\, 3\}} \, E_{h}\{\log
g^{w_2}(X; {\hat{\gamma}(j)} )\}~\hbox{v.s.}~ {\cal H}^{g^{w_2}}_{12}:Reject~{\cal H}^{g^{w_2}}_{02};\\
\nonumber {\cal H}^{g^{w_2}}_{03}: E_{h}\{\log
g^{w_2}(X;{\hat{\gamma}(3)} )\} \geq \max_{j \in \{1, \,2\}} \,
E_{h}\{\log g^{w_2}(X; {\hat{\gamma}(j)})\}~\hbox{v.s.}~ {\cal
H}^{g^{w_2}}_{13}: Reject~{\cal H}^{g^{w_2}}_{03}.
\end{eqnarray*}
Now, one may build up a $975\%$ model confidence set for the
second part. Table 4 shows decision of the above hypothesis tests
at significance level $\beta=0.025$ after 1000 irritations.
\begin{center}
\scriptsize{Table 4: Hypothesis tests at significant level
$\beta=0.025$ for
$A=(0,\infty)$} \tiny{\\
\begin{tabular}{c c c}\hline
Hypothesis Test & Test Statistics& Conclusion\\
\hline ${\cal H}^{g^{w_2}}_{01}$ v.s.  ${\cal H}^{g^{w_2}}_{11}$& $-1.540$& ${\cal H}^{g^{w_2}}_{01}$ is accepted\\
       ${\cal H}^{g^{w_2}}_{02}$ v.s.  ${\cal H}^{g^{w_2}}_{12}$& $-1.520$& ${\cal H}^{g^{w_2}}_{02}$ is accepted\\
       ${\cal H}^{g^{w_2}}_{03}$ v.s.  ${\cal H}^{g^{w_2}}_{13}$& $-3.360$& ${\cal H}^{g^{w_2}}_{03}$ is rejected \\
\hline
\end{tabular}}
\end{center}

Therefore, the desired $975\%$ local confidence set for the second
part is
\begin{eqnarray*}
\tau^{w_2}_2=\left\{ {\cal G}^{w_2}_{\Gamma(1)}; {\cal
G}^{w_2}_{\Gamma(2)} \right\}.
\end{eqnarray*}
    \item[Step 3: Mixture confidence set.] Now to construct a mixture $95\%$ model confidence set for unknown
density function $h(\cdot),$ we minimized ${\cal KL}$ distance
between convex combination of elements of $\tau^{w_1}_1$ and
$\tau^{w_2}_2$ and $h(\cdot).$  Such convex combination is given
by \footnotesize{\begin{eqnarray*}
  \tau &=& \displaystyle\bigcup_{(i,j)\in \{3, 4\}\times \{1, 2\}}\left\{\alpha^{opt}_{ij}{f^{w_1}}(x; {{\hat\theta}(i)} )+(1-\alpha^{opt}_{ij}){g^{w_2}}(x; {{\hat\gamma}(j)}):~~{f^{w_1}}(x;{{\hat\theta}(i)})\in
\tau^{w_1}_{1}~\&~{g^{w_2}}(x;{{\hat\gamma}(j)}) \in
\tau^{w_2}_{2} \right\}.
\end{eqnarray*}}\normalsize
Using result of Theorem \eqref{confidence_set_2-mixture}, we
estimate optimal mixture weight as $\alpha^{opt}_{31}=0.336$,
$\alpha^{opt}_{32}=0.335$, $\alpha^{opt}_{41}=0.337$ and
$\alpha^{opt}_{42}=0.335$. Table 5 represents the Hellinger and
the $L^{2}$ distances for the  $95\%$ mixture model confidence
set.

\begin{center}
\scriptsize{Table 5: The values of Hellinger and $L^{2}$ distances for the  $95\%$ mixture model confidence set} \tiny{\\
\begin{tabular}{c c c}\hline
Combining Models & Hellinger Distance& $L^{2}$ Distance\\
\hline $\alpha^{opt}_{31}{f^{w_1}}(x; {{\hat\theta}(3)})+(1-\alpha^{opt}_{31}){g^{w_2}}(x; {{\hat\gamma}(1)} )$& $0.0095$& $0.0052$\\
       $\alpha^{opt}_{32}{f^{w_1}}(x;{{\hat\theta}(3)} )+(1-\alpha^{opt}_{32}){g^{w_2}}(x; {{\hat\gamma}(2)})$& $0.0056$& $0.0035$\\
       $\alpha^{opt}_{41}{f^{w_1}}(x;{{\hat\theta}(4)} )+(1-\alpha^{opt}_{41}){g^{w_2}}(x;{{\hat\gamma}(1)} )$& $0.0080$& $0.0037$\\
       $\alpha^{opt}_{42}{f^{w_1}}(x;{{\hat\theta}(4)} )+(1-\alpha^{opt}_{42}){g^{w_2}}(x; {{\hat\gamma}(2)})$& $0.0043$& $0.0022$\\
\hline
\end{tabular}}
\end{center}
As Table 5 shows that all elements of the  $95\%$ mixture model
confidence set are appropriate choice for underling density
function $h(\cdot),$ given by Equation
\eqref{True-density-Example-2}.
\end{description}
Part (b) of Figure 1 illustrates element of the above $95\%$
mixture model confidence set and true density function. From this
figure and Table 5, one may conclude that the above $95\%$ mixture
model confidence set provides an appropriate approximation for
true density function \eqref{True-density-Example-2}.
\begin{center}
\begin{figure}[h!]
\centering\subfigure[]{
\includegraphics[width=18cm,height=9cm]{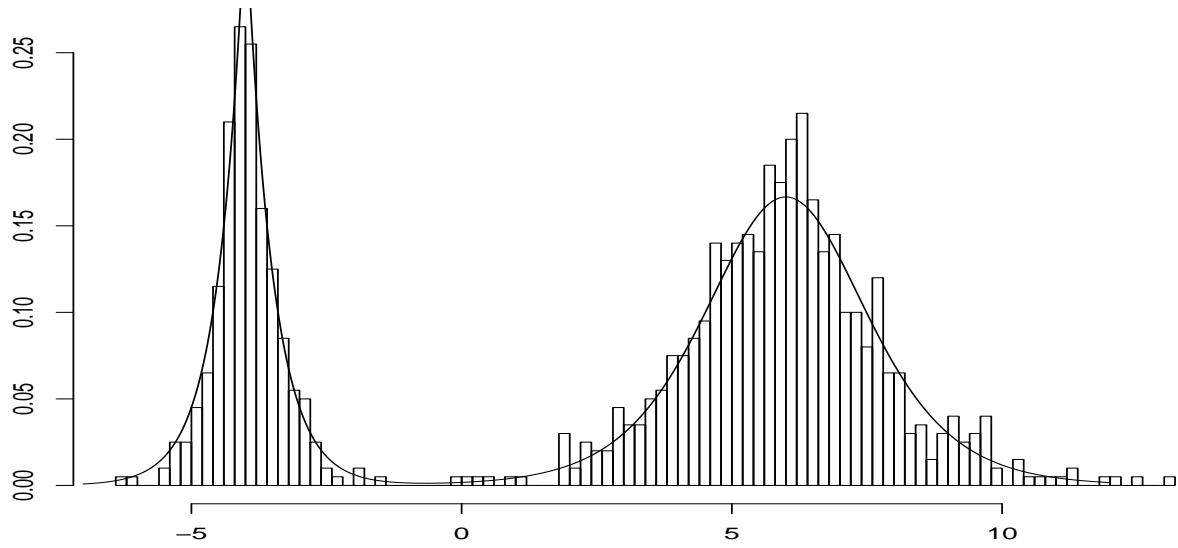}}\\\subfigure[]{
\includegraphics[width=18cm,height=9cm]{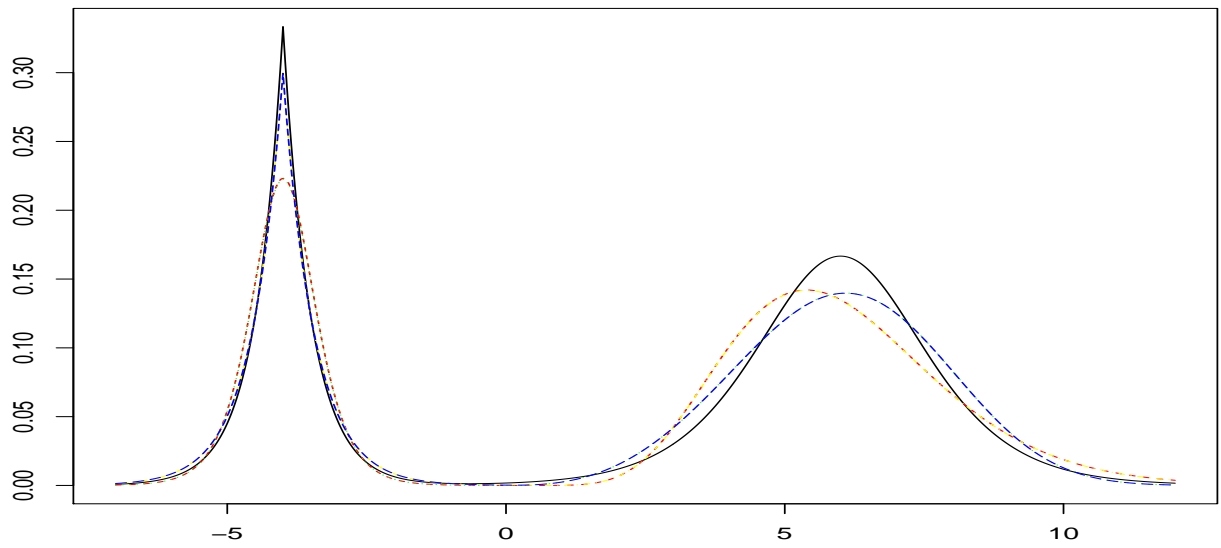}}
\caption{\scriptsize Part (a): Graphical illustration 1000
generated samples from true density function
\eqref{True-density-Example-2}. Part (b): Graphical illustration
of elements of the above $95\%$ mixture model confidence set and
true density function.}
\end{figure}
\end{center}
\section{Conclusion and Suggestions}
This article considers the problem of constructing an appropriate
model confidence set, whenever underling model has been
misspecified and some part of support of random variable $X$
conveys some important information above misspecified density
function $h(\cdot).$ Using weighted density functions, this
article constructs a weighted model confidence set for true
density function $h(\cdot).$ Applications for such weighted model
confidence set for local and mixture model confidence sets have
been given. Through a simulation study, we have been seen that the
weighted  model confidence set offers a convenient  model
confidence set for complex data. Our findings cannot practically
employ whenever non-nested competing model contain a large number
models. Therefore, we suggest to develop an one single procedure
to built up such confidence model set.
\section*{Acknowledgements}
Thanks to an anonymous reviewer for his/her constructive comments.

\end{document}